# A Case for 3D Integrated System Design for Neuromorphic Computing & AI Applications

Eren Kurshan

*Columbia University*
*New York, NY, 10027*
*ek2925@columbia.edu*

Hai Li

*Duke University*
*Durham, NC, 27701*
*hai.li@duke.edu*

Mingoo Seok

*Columbia University*
*New York, NY, 10027*
*ms4415@columbia.edu*

Yuan Xie

*U.C. Santa Barbara*
*Santa Barbaca, CA, 93106*
*yuanxie@ece.ucsb.edu*

Over the last decade, artificial intelligence has found many applications areas in the society. As AI solutions have become more sophistication and the use cases grew, they highlighted the need to address performance and energy efficiency challenges faced during the implementation process. To address these challenges, there has been growing interest in neuromorphic chips. Neuromorphic computing relies on non von Neumann architectures as well as novel devices, circuits and manufacturing technologies to mimic the human brain. Among such technologies, 3D integration is an important enabler for AI hardware and the continuation of the scaling laws. In this paper, we overview the unique opportunities 3D integration provides in neuromorphic chip design, discuss the emerging opportunities in next generation neuromorphic architectures and review the obstacles. Neuromorphic architectures, which relied on the brain for inspiration and emulation purposes, face grand challenges due to the limited understanding of the functionality and the architecture of the human brain. Yet, high-levels of investments are dedicated to develop neuromorphic chips. We argue that 3D integration not only provides strategic advantages to the cost-effective and flexible design of neuromorphic chips, it may provide design flexibility in incorporating advanced capabilities to further benefits the designs in the future.

# 1. Introduction

AI has found a wide range of application areas in today's society, well beyond the traditional use cases. Emerging applications range from autonomous vehicles, to diagnostic medicine, digital assistants, trading systems, legal services and many internet-of-things devices. AI applications have unique characteristics that differentiates them from traditional software solutions. These characteristics such as their data intensive nature and the heavy reliance on unstructured data, cause challenges in the traditional processor architectures. The separation of computing and memory, in the state-of-the-art *von Neumann* architectures results in frequent data movement, which, in turn, hurts both the system performance and energy efficiency V. Sze (2017).

For a long time, the prevailing thinking in the AI community was that AI has little no connection to the underlying architecture and hardware. Even though the idea to separate the brain from its functions seems unreasonable from a biological perspective, hardware and enabling technologies were perceived as exogenous to the field of AI (as reflected in AI conferences and journals). Lately, the perception has changed, as the limitations of AI implementations on traditional commodity hardware presented numerous challenges, such as performance and energy efficiency. As an example, during the Jeopardy's finals match, IBM's Watson System (based on a cluster of 90 Power750 servers) consumed approximately 85,000 Watts and required special power delivery infrastructure to compete with the human contestants (that consumed 20 Watts range) Greenemeier (2013). This discrepancy highlighted the importance of hardware and novel architectures for the future of AI.

In response to such challenges, neuromorphic computing gained significant interest. Neuromorphic computing refers to a range of brain-inspired computer architectures, devices, and models, C. D. Schuman (2017) that have been researched since 1980s C. Mead (1989) (the building blocks have been researched since the 1960s). The recent interest has been driven by the inherent advantages it provides to AI and machine learning applications (such as deep learning). 3D integration is an enabling technology for neuromorphic computing, in mimicking the heavily interconnected and plastic architecture of the human-brain, while providing the ability to continue the scaling laws that decades long sustained performance improvement for computing systems.

Traditional architectures experience serious issues in the post Dennard's scaling era R. H. Dennard (1974) and the imminent end of the Moore's Law Moore (1965). The cost and complexity of manufacturing in the current 5nm and smaller nodes highlight the physical limits of scaling and the need for alternative technologies. 3D's strategic advantages in neuromorphic computing and AI include: (i) *Technological advantages* such as enabling the scaling laws at a low cost; (ii) *System-level advantages* by reducing the issues caused by the memory wall and integrating promising emerging technologies in the same stack; (iii) *Strategic design advantages* in dealing with the design and manufacturing risks and uncertainties in the system design; and (iv)*Emerging advantages and opportunities* in providing unique capabilities to better enable biologically inspired capabilities. This paper aims to provide a high-level overview of the 3D integration characteristics and discuss the potential opportunities they provide in neuromorphic computing. The paper is structured as follows:

Section 2 provides a high-level overview of 3D and neuromorphic computing; Section 3 discusses the advantages and emerging opportunities of 3D integration in next generation neuromorphic systems; Section 3.5 provides a high-level overview of challenges; finally Section 4 provides conclusion discussions and outlook.

## 2. Background

### 2.1. *3D Integration*

Vertical integration concepts were first introduced in late 60s and followed by the demonstration of chips in late 70s and early 80s K. Onoda (1969); Morihiro (2015). Monolithic and wafer-level integration have since been researched by a large number of studies edited by P. Garrou (2008), Bernstein (2007), M.Leong (2003). 3D integration relies on bonding multiple silicon device layers together, through techniques like wafer bonding with SiO2 (silicon dioxide) fusion bonding, Cu fusion bonding, micro-bumps (Pb/Sn), polymer adhesive bonding edited by P. Garrou (2008),C.Ko (2010). Alternative interconnect technologies have been used to connect the layers through micro-bumps, through-silicon-vias (made of copper, tungsten) Lau (2011).

At a high-level, 3D integration technologies fall into 3 main categories:
(i)*Monolithic 3D Integration:* As one of the original focus areas of research in the 80s, monolithic 3D refers to the process of building 3D stacks from the ground up from an original layer M. Vinet (2014).
(ii) *Die-to-Die Integration:* refers to the integration at die-level, where multiple dies are manufactured, aligned and bonded together. Despite having pre-bonding testing advantages, die-to-die integration has cost and alignment challenges Knickerbocker (2008).
(iii)*Wafer-to-Wafer Integration:* refers to the process of manufacturing 3D layers in different wafers (which are processed, aligned and bonded at wafer layer). Wafer level handling and alignment has made it more popular due to the process characteristics S. Koester (2008). In wafer-level integration, the lack of testing at the die-level may cause a faulty die to turn the corresponding 3D stack to be faulty.

In recent years, significant progress has been made on the historical challenges of 3D, such as through-silicon-via (TSV) size and pitch scaling, die and wafer alignment as well as yield improvements S.Iyer (2015). As shown in Figure 1 TSV improvements over time provide unique opportunities for neuromorphic architectures moving forward.

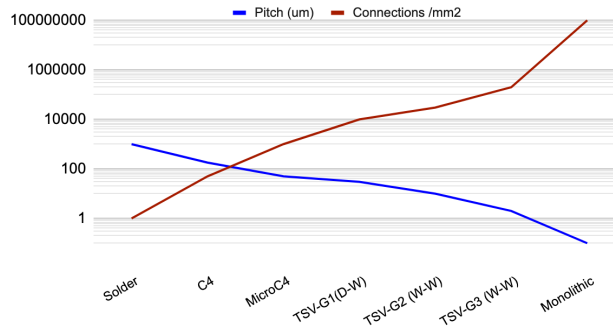

Fig. 1: 3D interconnectivity with different stacking technologies

## 2.2. *Neuromorphic Architectures*

Even though most neuromorphic architectures are loosely inspired by the human brain, they cover an extensive range of different styles. At a high-level, neuromorphic architectures fall into a spectrum of alternatives differentiated by their contrast to the traditional synchronous von Neumann architectures on one end and the distributed asynchronous systems on the other ORNL (2016).

(i)*Biologically Inspired:* Neuromorphic architectures span a spectrum in terms of biological inspiration, from loosely inspired chips, to the neural emulators that follow strict biological-plausibility guidelines.

(ii)*Application-driven:* For application-driven architectures, the design goal is to optimize the architecture based on the application characteristics (such as image and natural language processing, and other deep learning workloads R. Adolf (2016)).

However, as noted in Section 1, in recent years the applications of AI has significantly expanded, though the emerging applications share basic characteristics such as the large data sizes, heavy use of unstructured data, approximate computing for decision-making, adaptation to dynamic environment changes. As ORNL (2016) and R. Adolf (2016) showcase, neuromorphic applications further share characteristics like: noisy input data, multiple-modalities, not high-precision/approximate computing, spatiotemporal computations, the requirement of robustness, need for low power usage, continuous learning and real-time processing. In this paper, we try to argue both from the perspective of application drivers and biological inspirations.

## 3. 3D Neuromorphic System Drivers

3D integration provides a number of unique opportunities for neuromorphic computing systems. This section overviews the opportunities from technological, system-design, design strategy perspectives, in addition to highlighting emerging opportunities.

### 3.1. *Technological Opportunities*

**Opportunity 1.** 3D provides enhanced yield and reduced manufacturing costs in the post-Moore's Law era.

Despite the ever-growing interest in neuromorphic chips all semiconductor chips face serious challenges in the post-Dennard/post-Moore scaling era. Recent studies highlight the sharp increases in the chip manufacturing and IC design costs (including manufacturing, hardware verification, software, validation, prototyping, architecture stages). According to IBS, the design cost for 3nm chips are expected exceed 1 Billion U.S.D.. In addition, process development costs for 3nm were projected to be around $4 Billion to $5 Billion in process development (Lapedus (2019)). The fab cost for 40,000 wafers per month is projected to be in $15 Billion to $20 Billion range. These trends highlight a 70% cost increase from 10nm to 7nm and a 82% cost increase from 7nm to 5nm Lapedus (2018). Beyond 3nm, 2nm and smaller nodes introduce even higher levels of uncertainty and increases in the costs M. LaPedus (2020). 3D integration provides an alternative path for performance improvement by packing more transistors in the same stack P.Emma (2008).

Moreover, 3D enables yield and cost improvements over 2D by reducing the die sizes and pre-integration testing, which enables yield improvements. As 3D integration relies on system-level advantages, older and cheaper technology nodes may be used to achieve higher performance. According to S. Sinha (2020), below 10nm technology nodes these advantages may be prominent. Heteregeneous 3D stack of a 5nm and 7nm layers provides the cost benefit of 26% lower die-costs. Similarly, a 3D stack of 7nm layers yields 32% lower cost. In order to achieve these advantages 3D design tools are essential to effectively manage the trade-off (e.g. yield, performance improvement, testing overhead) Y.Chen (2010); Lau (2010).

### 3.2. *System-level Opportunities*

**Opportunity 2.** 3D allows full-system integration of all design components and chips in the high-density stacks.

***Application-level Drivers:*** In recent years, a number of neuromorophic chips have been developed: IBM's TrueNorth (A.S. Cassidy (2014)), Intel's Loihi (M.Davies (2018)), Spinnaker (M.M. Khan (2008); C. Liu (2018); S. Furber (2013)), BrainScales (Meier (2015)), DYNAPS (S. Moradi (2018)), NeuroGrid (B. V. Benjamin (2014)) etc. Most neuromorphic chips serve as building blocks in larger systems, typically connected at board-level. As an example, SpiNNaker system is constructed of 18 processor cores on a die, where 48 chips with 864 cores are joined on a board M.M. Khan (2008). Similarly, at the system-level 16 TrueNorth chips are joined on NS16e board A.S. Cassidy (2014).

3D integration provides the opportunity to integrate the full system in the same stack. This significantly reduces the inefficiencies of off-chip communications, limited bandwidth and interconnect delays. *Bandwidth wall* has been an important challenge in traditional processor architectures. Although the neuromorphic chips provides some advantages by integrating memory in the same stack with the processor, off-chip bandwidth is still an critical consideration for system-level performance B. Rogers ([n.d.]). I. K. Schuller (2015) reports approximately 20 times difference in packaging density between silicon-based computing and biological counterparts. This, combined with the increased interconnectivity, and other factors creates a significant difference in the resulting system-level performance.

**Opportunity 3.** 3D enables energy efficiency and lower power dissipation.

***Application-level Drivers:*** Energy efficiency and peak power dissipation have been considered among the primary drivers of neuromorphic computing. As discussed earlier in Section 1, power dissipation in deep learning applications is a primary concern. As an example, TrueNorth performs 46 billion synaptic operations per second (SOPS) per Watt. The corresponding energy dissipation is 26 pico-Joule per synaptic event with a power density of 20 mW/cm2. Similarly, Loihi delivers 30 billion synaptic operations per second, consuming about 15 pico-Joule per synaptic operation. Even though the power density is orders of magnitude smaller than standard CPUs, Truenorth A.S. Cassidy (2014) is still 10,000 times more power hungry compared to human brain I. K. Schuller (2015). Beyond the peak power or power density challenges, in Internet-of-Things systems neuromorphic computing and *in-situ* learning becomes more challenging due to energy efficiency and battery life concerns. 3D provides a few key advantages: (i) it enables the integration of more energy efficient device types and heterogeneous integration (ii) it reduces the energy inefficient data movements both on- and off-chip. As an example, according to M.M. Sabry Aly (2019) N3XT improves the energy efficiency of abundant-data applications 1000-fold by using new logic and memory technologies.

***Biological Drivers:*** Neuromorphic architectures employ a wide range of design choices to reduce the power dissipation and improve energy efficiency. For instance, they use spikes to implement event based computation, leverages communication systems (that consume energy only when needed), place synapses with memory to keep data movement local. They also limit connectivity to implement fan out efficiently to reduce the network traffic. Yet, comparing the machine learning implementations to biological systems (e.g. insect nervous systems) provides striking differences and inefficiencies Corrigan (2019). I. K. Schuller (2015) reports 100,000 times difference in the overall energy efficiency of the current computing systems and biological counterparts.

**Opportunity 4.** 3D enables enhanced local and global interconnectivity.

***Application-level Drivers:*** As noted earlier, traditional processor architectures face interconnectivity limitations such as on-chip bus bandwidth constraints, limited number of BEOL(back-end-of-the-line)/wiring layers, off-chip communication bottleneck. Increased interconnectivity and efficient movement of data within the system, are highly desired characteristics in AI applications. Neuromorphic architectures frequently use inter-chip P. Merolla (2014) and intra-chip NoC multi-cast techniques D. Vainbrand (2010). 3D integration provides enhanced connectivity in all directions (x,y,z), as well as significantly improved bandwidth and reduced memory and off-chip communication pressures.

Recent advances in the 3D manufacturing process(such as back-end-of-the-line/interconnect bonding with layers P. Batra (2014) as well as overlay tolerances in sub-micron range W. Lin (2014), S.Iyer (2015)) enables enhanced interconnectivity compared to earlier 3D chips. A. Kumar (2017) discusses the feasibility of 1 m vias, with an approximate pitch of 2-2.5m in high volume manufacturing environments, yielding a density of about *200,000 vias/mm2*. Wafer-scale integration has reached interconnectivity levels, where each node can also be connected to another node, fixed distance away both in the x-direction and in the y-direction and even in the z-direction A. Kumar (2017). Furthermore, recent 3D demonstrations provide even more promising results for monolithic (bottom up) process of manufacturing stacked layers S.K.Moore (2019).

***Biological Drivers:*** Human brain exhibits high-degrees of interconnectivity, both globally across different brain regions and locally within individual regions P.Hagmann (2003). According to I. K. Schuller (2015), a comparison of neuromorphic architectures and biological systems reveal the major differences in fan-out connectivity by approximate 10,000 times. Similarly, the ratio of synapses to neurons in current neuromorphic architectures range from (244 in TrueNorth A.S. Cassidy (2014) to 1000 in SpiNNaker (M.M. Khan (2008); C. Liu (2018))), while in the human brain this ratio is 10,000 M.A. Ehsan (2017). 3D integration has the potential to provide significant improvement in the neuron density and interconnectivity in neuromorphic architectures.

Beyond the high-degree of connectivity, R. Singh (2015) highlights a significant amount of diversity in the local and global interconnectivity patterns in primate brains. At the local-level, various types of connectivity patterns exist, such as star-like or clique-like networks. However, such diversity of connectivity does not exist in current neuromorphic architectures due to the limitations of the wiring-layer connectivity in 2D chips. 3D integration enables the implementation of a diverse range of interconnect patterns locally.

**Opportunity 5.** Heterogeneous 3D allows the integration of disparate technologies and diverse building blocks.

***Biological Drivers:*** Korr (1980) highlights that human brain has over 100 different types of cells, neurons, glia (astrocytes, oligodendrocytes), epythelial cells, microglia, ependymal cells Lake (2016). Structurally, the brain has neurons and neural networks variations in different regions (such as pyramidal neurons in the cortex, basket cells, spindle cells for interconnectivity, Purkinje cells). Each neuron type exhibits unique connectivity and architectural characteristics. However, this diversity is difficult to implement in current neuromorphic architectures and manufacturing processes. 3D provides a possible path to integrate different neuron types, device and connectivity patterns.

***3D Advantages:*** 3D provides considerable opportunities in integrating disparate technologies seamlessly, hence building heterogeneous systems with different components (sensors, memory, computing, communication macros) as well as disparate technologies (such as PCM, RRAMs, different technology nodes, analog/digital designs) in the same stack C. D. Schuman (2017).

Different technology nodes (older technology nodes of 5-7nm to higher), DRAM memory and non-volatile storage layers, programmable arrays (e.g. field programmable gate arrays (FPGAs) M. Banuelos-Saucedo (2003), analog arrays (FPAAs) P. Rocke (2008), neural arrays (FPNAs) E. Farquhar (2006), mixed analog and digital designs can be integrated. ASICs have been among the most historical implementations of neural networks. 3D integration enables ASIC components F. Blayo (1989); F. Distante (1990) along with standard functional blocks and communication modules. A wide range of device types including memristors G. C. Adam (2017); H.An (2018); G. Indiveria (2013), non-volatile memory (such as ReRAMs for memory P. Chi (2016), phase change memories C. D. Wright (2013) to implement synapses), spintronic devices K. Roy (2014), transistor architectures (such as floating gate transistors S. Aunet (2003) as storage or amplifiers or neuron components), atomic switches M.Suri (2012) can be integrated. Similarly, a range of materials from graphene H. Tian (2015), to carbon nano-tubes (CNT) C. Chen (2012) and other emerging technologies have been successfully integrated as parts of 3D stacks. Heterogeneous integration of a selection of the technologies have already been manufactured and demonstrated (e.g. processor/DRAM stacks, CNT/ReRAM stacks S.K.Moore (2019)).

**Opportunity 6.** 3D integration provides performance improvement over 2D baseline alternatives through system-level design.

Over the years, 3D architecture research focused heavily on multi-layer memory modules as well as processor/memory integration for traditional multi-threaded chip multi-processors. In neuromorphic computing more sophisticated stacked architectures are possible.
***3D Advantages:*** P.Emma (2008) and S. Sinha (2020) argue that for traditional architectures 3D provides significant performance improvement compared to manufacturing in advanced technology nodes. According to S. Sinha (2020), 3D mapping of a conventional 2D design translates to performance improvement of up to 12% or power reduction of up to 40% X.Xu (2019). It also highlights significant improvement in architectural performance metrics such as instructions per cycle (IPC) due to the improved cache hierarchy and increased cache capacity. Recently, monolithic 3D systems have been demonstrated. Among these the nano-engineered computing systems technology M.M. Sabry Aly (2019) integrates all systems components in the same die and removes the off-chip memory hierarchy access (which improves the performance significantly) and claims to achieve orders of magnitude improvement in energy efficiency by using a range of technologies and heterogeneous integration.

**Opportunity 7.** 3D has the ability to incorporate structural reconfigurability in neuromorphic architectures.

***Biological Drivers:*** Plasticity and reconfigurability, by design, are key characteristic of the human brainL. J. Hearne (2017). Brain reconfiguration spans an extensive list of complex processes of different kinds and levels such as: (i) Synaptic connectivity (ii) Mode dependent states (during different states of the brain L. J. Hearne (2017)), (iii) Functional reconfigurability for resilience after damage or removal of parts of the brain (such as hemispherectomy V. Holloway (2001)).

***3D Advantages:*** 3D provides unique opportunities in reconfigurability by integrating configurable structural blocks at different levels of granularity, such as programmable gate arrays, atomic switches, reconfigurable on-chip interconnect solutions. As an example, nFPGA architecture utilizes 3D integration techniques and new nanoscale materials for stacked FPGA implementation C. Dong ([n.d.]). J.Cong (2009) proposes FPGA architecture with memristor-based reconfiguration (mrFPGA), which is also CMOS compatible. Programmable interconnects of mrFPGA use memristors and metal wires, where interconnects are fabricated over logic blocks providing higher connectivity and lower footprint. 3D-aware placement of the macros further optimizes the resulting reconfigurable architectures.

**Opportunity 8.** 3D facilitates modular integration of layers, functionalities, technologies and system components, hence the ability to construct alternative neuromorphic systems at a low cost.

***Biological Drivers:*** While it cannot be interpreted as intentional, human brain has been built in a modular fashion such that advanced functions are incorporated over more primitive brain structures (from the reptilian brain in brain stem and cellebrium, mammalian brain in limbic system, and human brain in neo-cortex) Rakic (2009). Even though neuromorphic architectures, by design, have a completely different design evolution, the functional separation and integration of different capabilities is of great interest.

***Application Drivers:*** 3D provides a platform to make long-term or short-term changes to the structure of the neuromorphic architectures, without causing costly redesign exercises for the entire system. As long as the infrastructure components such as power and signal connections for future layers are accounted for and specified, novel design layers can be integrated with minimal cost overhead. This serves multiple purposes: (i) Gradual additions to the design to deal with evolving application demands or design changes; (ii) Customization of the system architecture for different applications (e.g. IoT, embedded systems, or high-end use cases); (iii)

Biologically-inspired functionalities enabled by the modular architecture, such as interactions between different layers (such as control, master-slave functionality between lower and higher functionalities neocortex and other brain regions); (iv) Repairs and redundancies for robustness purposes.

***3D Advantages:*** While modularity provides limited design time flexibility for bottom-up monolithic 3D stacks, wafer-scale or die-level integration enables flexibility in the later stages, such that: (i) each layer and die can be manufactured and tested separately with disparate manufacturing processes; (ii) with the use of standardized interfaces and infrastructure components individual layers can be switched with alternatives in the later stages. For example, as long as the signal and power delivery placement is standardized, different technologies can be integrated in the stack later in the design cycle. P.Emma (2010) argues that the characteristics of 3D integration enables customization of the system design for a range of architectures, based on use cases, application demands and other criteria. Customized neuromorphic stacks can be manufactured at a lower cost for different use cases and applications.

**Opportunity 9.** 3D provides size and form factor advantages for embedded and IoT applications.

By vertically integrating the layers, 3D reduces the 2D chip footprint close to proportional to the number of layers. Through-silicon-vias for signal and power interconnectivity, keep-out zones as well as placement complexities create overhead, which can be significant for larger via sizes and high-power density stacks. However, according to recent studies S.K. Samal (2016) for via sizes of 2m, TSV area overhead can be below 20% even for traditional architectures. Lower form factor is a significant advantage in IoT applications and embedded systems, in which larger die sizes have cost, yield, size, timing, and performance implications.

### 3.3. Design Strategy

**Opportunity 10.** The strategic design flexibility provided by 3D is critically important to tackle the unknowns in the biological systems, as well as the uncertainty in the resulting designs and emerging technologies.

***System Design Drivers:*** Despite the historical roots in 1980s neuromorphic computing is in early stages of wide-scale adoption. A large amount of uncertainty exists due to the dueling forces of:

(i) *Uncertainty in AI and deep learning applications:* Number of researchers expressed concerns on the strategic direction of AI and deep learning solutions. Recently Lecun posed the question Lecun (2019) *"Why build chips for algorithms that don't work?"*. This becomes a concern for AI and deep learning hardware as chip design requires considerably higher investments.

(ii) *Limited understanding of the human brain:* Although, significant progress has been made in recent years, the architecture and functionality of the human brain remains a mystery. Similar to Lecun's question on the applications, neuromorphic computing faces another fundamental challenge - the question of *"How to build chips to emulate biological systems we don't understand"*.

Meanwhile, brain-inspired neuromorphic chips are developed in full force, studies such as S. Jäkel (2017) highlight how the current understanding of the function of human brain is limited and may be incorrect. S. Jäkel (2017) posits that neuron and synapse based models of the brain may be missing a large part of the brains functionality by disregarding glial functions. Understanding the brain functionality at and beyond the connectome-level will likely take time. Meanwhile, neuromorphic chips require significant investment and time despite the uncertainty in their design goals. Similarly, from a scaling perspective, citeBengioLecun outline the challenging road ahead in scaling the current deep learning kernels and network architectures to the target ranges.

***3D Advantages:*** 3D integration provides a platform to hedge the potential risks of architectural and technological risks and uncertainties during the earlier stages of neuromorphic computing. Due to the cost and complexity advantages, perhaps one of the most important motivator of 3D neuromorphic is this 3D risk mitigation. The current ambiguity of AI and deep learning solutions motivate lower cost and flexible design frameworks, able to adopt to changes in the technological advances in the future. 3D integration provides such a platform, which can leverage investments into a diverse set of design paradigms and technologies. During the early stages of the design process, as the selected system design commits to a specific type of technology it also frequently restricts itself and the design-space (from using incompatible technologies). Over time, as technologies mature, such restrictions may be costly to change and revise as they frequently require going back to earlier design stages. The high costs for design and manufacturing hence motivates the strategic design flexibility 3D integration provides Y. Xie (2006).

### 3.4. *Emerging Opportunities*

**Opportunity 11.** 3D has intrinsic advantages in the construction of biologically-inspired architectures and distinct layers of functional specialization.

***Biological Drivers:*** As discussed in the earlier sections, human brain and other biological systems have notable differences in their characteristics such as: higher interconnectivity, higher neuron density, lower energy dissipation. 3D provides unique capabilities in bridging the gap between the biological systems and state-of-the-art semiconductor chips. Exotic technologies, devices and materials needed for neuromorphic computing may not always be compatible in standard silicon processes, heterogeneous integration of disparate technologies in the same stack enables the use of such technologies.

***3D Advantages:*** In recent years, neuro-inspired architectures and AI solutions have exhibited promising results (such as in generalizations, scene understanding use cases) D. Hassabis (2017). In a 3D stack, different functional modules and layers may be dedicated to distinct capabilities such as image processing, speech processing, face-recognition, spatial navigation. This provides the ability to incorporate disparate functionalities in a fully integrated system design. Customized designs can be constructed by varying the use of functional layers for application-specific requirements (e.g. image processing layers used in one IoT stack may not be needed in another).

**Opportunity 12.** 3D integration of emerging device types and manufacturing technologies have the potential to better enable online and in-situ learning.

Software-based deep learning implementations frequently require a large number of servers and computing resources for compute intensive training stages (especially in back-propagation based approaches). In contrast human brain and biological systems learn online, in-situ and real-time. Even though online learning is seen as the future in AI community, traditional computing systems are not amenable to implement on-line learning efficiently Lecun (2019). Neuromorphic architectures provide the ability to mimic the biological processes in neurons C. D. Schuman (2017) through electro-chemical processes in artificial neuron implementations such as learning through tunable resistance of the memristor arrays C. Li (2018), synaptic learning mechanisms Z. Q.Wang (2012). These core capabilities play a key role in neuromorphic architectures and motivate heterogeneous integration of memristor and other enabling device layers in heterogeneous 3D stacks.

**Opportunity 13.** 3D can be used to implement dynamic controls and modulation.

***Biological Drivers:*** Human brain has numerous modulation and control mechanisms that are largely missing in the current neuromorphic architectures ORNL (2016). Among these, neurotransmitter based modulation, limbic and emotional modulation of the brain activity are just a few. As an example, recent studies have highlighted biological mechanisms such as the local control and modulation of the blood flow through glial cells and neurons D. Attwell (2010). Even though current neuromorphic architectures do not have such efficiency and modulation mechanisms, they can be achieved through dedicated 3D vertical power delivery systems placed along with control and support at the substrate layer. At a higher level, brain functionality is also modulated through limbic system. Depending on the environmental conditions, parts of the brain may be over or under activated for speed and the effectiveness of response. As an example, during the *fight or flight* response, the human brain is modulated through the limbic system and hormones such as cortisol S.Vogel (2016). Cortisol blocks hippocampus as well as other memory related regions of the brain while activating others to enable (or disable) the different regions necessary (or unnecessary) for survival. Other hormones and glucocorticoids are reported to play an important role on the human cognition and learning Wirth (2015).

***3D Advantages:*** Brain-like sophisticated modulation and control is far from reality in the current neuromorphic systems. 3D integration provides numerous knobs through electrochemical processes, that can serve as controller mechanisms for different layers, technologies and regions. Threshold voltage plays an important role in some designs such as M.M. Khan (2008). 3D substrate control has the ability to enable fine tuning and modulation of such neuromorphic chips. Similarly, such controller mechanisms can enable fine-grain or global modulation of the different functional regions.

**Opportunity 14.** Enhanced robustness and adaptability

***Biological Drivers:*** Human brain and biological neural systems have intrinsic redundancy in their structures R.B.Glassman (1987). They also exhibit high levels of adaptability and robustness. Current neuromorphic architectures lack functional specialization similar to biological systems (e.g. visual cortex). Hence it is difficult to argue for redundancy in a non-specialized homogeneous architectures. However, the lack of specialization also has significant impact on the performance and efficiency of the neuromorphic architectures.

***3D Advantages:*** Within the trade-offs of functional specialization and redundancy for robustness, 3D integration provides the ability to integrate redundancy in the system architecture without causing significant impact on the cost, performance or power density (while maintaining high degrees of functional specialization). Reconfiguring, switching on/off column level redundancies, modulating layers similar to human brain is a possibility by physical separation of the device layers and can be further optimized through enhanced placement and specialized control mechanisms.

### 3.5. *Implementation Considerations*

3D integration requires design considerations to tackle the vertical integration complexities K. Bernstein (2007).

**Consideration 1.** Power delivery, heat removal and management are still important challenges in 3D.

Traditional architectures faced unique challenges due to their high power consumption and power density in 3D E. Kursun (2010); E.Kursun (2012); Q.Zou (2017). Neuromorphic architectures are orders of magnitude lower power compared to the von Neumann counterparts. This is an intrinsic advantage in power delivery and thermal management of the resulting chips. In some of the latest 3D implementations, the device layers are thinned down to 5m range with up to 4 layers of silicon A. Kumar (2017). Thinned silicon layers have been shown to exacerbate peak temperatures due to the lack of heat flow in the bonding and BEOL layers E. Kursun (2010), micro-channels and other cooling solutions may be challenging to implement in advanced 3D stacks due to the layer thicknesses A. Kumar (2017). In such many-layer 3D stacks, bonding materials and BEOL layers (with different thermal characteristics than silicon layers) E.Kursun (2012), macro placement and activity management require special design considerations. Similarly, power delivery for heterogeneous 3D stacks require tool and infrastructure support. Recent neuroscience studies indicate that human brain has specialized biological mechanisms for heat removal, temperature modulation and control H.Wang (2016).

**Consideration 2.** Limited availability of design tools and methodologies for neuromorphic 3D remains as a challenge.

In recent years, a number of research studies have explored partitioning, placement, timing, routing and other aspects of 3D design process C.Chiang (2009); J.Cong (2009). Furthermore, more integrated design tools and solutions have been developed. However, industry-grade tools are still limited and are critically needed. Most of the current industrial tool rely on having each 3D tier to be designed and optimized separately S. Sinha (2020), which causes numerous challenges. Similarly, a level of standardization in power delivery and communication infrastructures is needed to better utilize the modular integration potential and reduce the cost of building a diverse range of layer alternatives.

## 4. Conclusions and Outlook

Over the past decade, as AI has expanded its application range, it underscored the weaknesses of traditional von Neumann architectures. In response, neuromorphic architectures have gained significant popularity. Neuromorphic architectures provide intrinsic advantages over traditional chip multi-cores and have been demonstrated to yield higher performance and lower power in the emerging AI and deep learning use cases. 3D integration provides a number of key advantages in neuromorphic system design. These advantages range from technological advantages, to system-design opportunities, application level advantages and strategic advantages in designing and manufacturing costly hardware systems. (i) In addition to providing key advantages in performance, cost, interconnectivity, power efficiency, modularity, form factor; 3D integration is unique in its ability to enable the use of a diverse set of architectures, devices and technologies for neuromorphic computing. (ii) Biological brains showcase the full integration of memory, computing and communication capabilities. 3D integration provides the opportunity to integrate all 3 functions that may be implemented in disparate technologies or design components. (iii) Finally, emerging technology development typically involves large investments (financially, in addition to time and resources). In neuromorphic computing the risks on such investments are higher due to the broader uncertainties of the applications and the enabling technologies. The road map from the current neural network kernels to full fledged architectures to support general intelligence is considered a challenging and uncertain path. 3D provides a path to hedge these risks to a certain extent, while providing strategic design flexibility during the development and manufacturing processes. Even though the lack of understanding of the human brain will likely continue driving the architectural uncertainty in neuromorphic computing in the near future, in the post-Dennard/post-Moore scaling era, 3D integration can provide the critically needed capabilities towards the development of next-generation neuromorphic chips.